\documentclass[preprint,prd,amsmath,amssymb,amsthm,nofootinbib]{revtex4}
\usepackage{amsmath}
\usepackage{amscd}
\usepackage{amssymb}
\usepackage{amsfonts}
\usepackage[mathscr]{euscript}
\usepackage{ulem}
\usepackage{bm}
\usepackage{graphicx}
\usepackage{subfigure}
\usepackage{multirow}
\usepackage{amsmath,amssymb,amsthm}
\usepackage[colorlinks=true,linkcolor=red]{hyperref}
\newcommand{\xmark}

\newcommand{\bea}{\begin{eqnarray}}
\newcommand{\eea}{\end{eqnarray}}
\newcommand{\beq}{\begin{equation}}
\newcommand{\eeq}{\end{equation}}

\def\/{\over}

\begin{document}

\title{Quantum thermal field fluctuation induced corrections to the interaction between two ground-state atoms}

\author{Shijing Cheng$^{1,2}$, Wenting Zhou$^{3,}$\footnote{Corresponding author: zhouwenting@nbu.edu.cn}, and Hongwei Yu$^{4,}$\footnote{Corresponding author: hwyu@hunnu.edu.cn }}
\affiliation{$^{1}$School of Fundamental Physics and Mathematical Sciences, Hangzhou Institute for Advanced Study, UCAS, Hangzhou 310024, China\\
$^{2}$School of Physical Sciences, University of Chinese Academy of Sciences, No.19A Yuquan Road, Beijing 100049, China\\
$^{3}$Department of Physics, School of Physical Science and Technology, Ningbo University, Ningbo, Zhejiang 315211, China\\
$^{4}$
Department of Physics and Synergetic Innovation Center for Quantum Effect and Applications,
Hunan Normal University, Changsha, Hunan 410081, China
}

\begin{abstract}
We generalize the formalism proposed by Dalibard, Dupont-Roc, and Cohen-Tannoudji [the DDC formalism] in the fourth order for two atoms in interaction with scalar fields in vacuum to a thermal bath at finite temperature $T$, and then calculate the interatomic interaction energy of two ground-state atoms separately in terms of the contributions of thermal fluctuations and  the radiation reaction of the atoms and analyze in detail the  thermal corrections to the van der Waals and Casimir-Polder interactions. We discover  a particular region, i.e.,  $\sqrt[4]{\lambda^3\beta}\ll L\ll \lambda$ with $L$, $\beta$ and $\lambda$ denoting the interatomic separation, the wavelength of  thermal  photons and the transition wavelength of the atoms respectively, where the thermal corrections remarkably render  the van der Waals force, which is usually attractive, repulsive, leading to an interesting crossover phenomenon of the interatomic interaction from attractive to repulsive as the temperature increases.  We also find that the thermal corrections cause significant changes to  the Casimir-Polder force when the temperature is sufficiently high,  resulting in  an attractive  force proportional to $TL^{-3}$ in  the $\lambda\ll\beta\ll L$ region, and a force which can be either attractive or repulsive and even vanishing  in the $ \beta\ll\lambda\ll L$  region depending on the interatomic separation.

\end{abstract}

\maketitle

\section{Introduction}

One of the remarkable consequences of quantization of electromagnetic fields is the existence of vacuum fluctuations as a result of the Heisenberg uncertainty principle, which has profoundly changed our usual understanding of a vacuum.
The vacuum fluctuations are deemed as the origin of various physical phenomena, such as the spontaneous emission~\cite{Milonni76} and the Lamb shift~\cite{Lamb47},
the attractive force between two perfectly conducting neutral plates in vacuum, which is  known as the Casimir effect~\cite{Casimir48}, as well as
a  force felt by a neutral atom near a perfectly conducting plate, due to the coupling between the atom and the vacuum electromagnetic fluctuations modified by the presence of the boundary~\cite{Polder48}.

When two atoms are involved, a  vacuum-fluctuation-induced interatomic interaction then appears, which can be physically understood as follows.  One atom interacts with  fluctuating vacuum electromagnetic fields, and a radiative field is subsequently produced, which then acts on the other atom, and vice versa. As a result, an interaction potential, which is also generally referred to as the van der Waals or Casimir-Polder interaction potential, is induced~\cite{London30,Polder48,Craig98,McLone65,Holstein01,Power93,Power95,Salam10}.
Based on the standard perturbation theory, London  found that, for two ground-state atoms with an interatomic distance $L$ much smaller than the characteristic transition wavelength of the atoms, the interatomic interaction potential 
behaves as $L^{-6}$~\cite{London30}, and Casimir and Polder further showed that in the far region where the interatomic distance $L$ is much larger than the atomic transition wavelength, the  potential decays more rapidly  with a behavior of $L^{-7}$ rather than $L^{-6}$, due to the effect of retardation~\cite{Polder48}.
A few years later, Lifshitz {\it et al.} developed a macroscopic theory for the interaction of dielectric bodies, and by going to the limiting case of rarefied media, they recovered both the nonretarded and retarded interatomic interaction potential~\cite{Lifshitz56,Dzyaloshinskii61}. 

The early works mentioned above  are about the interatomic interaction energy of two ground-state inertial atoms coupled to the fluctuating electromagnetic field in a trivial flat space in vacuum. Over the past decades, endeavors have also been devoted to this fluctuation-induced interaction in more complex circumstances, such as atoms in a bounded space~\cite{Passante07,Donaire17}, in noninertial motions~\cite{Noto13,Noto14}, in a thermal bath~\cite{McLachan63,Boyer75,Milonni96,Ninham98,Goedecke99,Wennerstrom99,Passante07,Sherkunov09}, and even in a curved spacetime~\cite{Zhang11,Menezes17}.
For the effects of a thermal bath, let us note that after the work of Lifshitz and his collaborators~\cite{Lifshitz56,Dzyaloshinskii61}, McLachlan revisited the long-range dispersion forces in a non-uniform dielectric at finite temperature~\cite{McLachan63}.
He derived most of the results of Lifshitz {\it et al.} in a more elementary way, and discovered that the interaction energy of two atoms in a thermal bath at a low temperature $T$, which is much smaller than the energy gap of the atoms $\omega_0$, scales as $L^{-6}$ at small interatomic distances and as $L^{-7}$ at large interatomic distances when the retardation effect is important;
while at sufficiently long distances, $L^{-1}\ll T\ll\omega_0$, the temperature effects mask the retardation effect and the interatomic interaction energy scales as $TL^{-6}$~\cite{McLachan63}.
Later, this temperature  and distance dependence was also shown by Boyer in the theory of classical electrodynamics with classical electromagnetic zero-point radiation~\cite{Boyer75}, by Milonni and others via considering the energy of an induced dipole moment of a molecule in an electric field correlated with the second molecule~\cite{Milonni96,Goedecke99}, 
and by Wennerstr$\mathrm{\ddot{o}}$m and his collaborators with two different methods [the Lifshitz theory and the generalized fourth-order quantum-electrodynamical perturbation theory]~\cite{Wennerstrom99}. In addition, by taking the dilute medium limit of the Lifshitz theory~\cite{Lifshitz56,Dzyaloshinskii61} for two parallel dielectric slabs in a thermal environment, Ninham and Daicic also recovered the London potential in the nonretarded region and revealed that as the retarded limit is approached with increasing interatomic separation, temperature effects become increasingly significant and ultimately the London description must break down, even at a fixed temperature~\cite{Ninham98}. Recently, the study of  the thermal effects on the dispersion interaction energy of two ground-state atoms has also been extended to the case of atoms out of thermal equilibrium~\cite{Sherkunov09}.

When a two-atom system in a thermal bath is concerned,   there are three characteristic physical  parameters, i.e., the interatomic separation, $L$, the transition wavelength of the atoms, $\lambda\sim\omega_0^{-1}$, and the wavelength of thermal photons, $\beta=T^{-1}$, and so there are in general six typical regions for the problem under consideration. However, the previous studies on this topic are only concerned with one or a few regions, and to the best of our knowledge, a complete analysis of the interatomic interaction at finite temperature is still lacking. In this paper, we aim to fill this gap, i.e.,   we will  give a complete study for the interaction potential of two ground-state atoms in a thermal bath at an arbitrary temperature.

We exploit the formalism proposed by Dalibard, Dupont-Roc, and Cohen-Tannoudji [the DDC formalism]~\cite{DDC82,DDC84}, which separates the contributions of field fluctuations and the contributions of radiation reaction of atoms, to deal with the interaction potential. It is worth pointing out that the DDC formalism has been widely used, during the past years,  to study  various second-order perturbation quantum effects, such as the energy shifts of a single atom~\cite{Audretsch95,Audretsch951,Passante98,Tomazelli03,Rizzuto07,Rizzuto091,Zhu10,Rizzuto11} and the resonance interaction energy of two entangled atoms~\cite{Rizzuto16,Zhou16,Zhou17,Zhou18,Zhou1897,18Zhou}, and very recently, it was generalized from the second order to the fourth order to study the interatomic interaction between two ground-state atoms in vacuum~\cite{Noto14,Menezes17,Zhou21}. In this paper, as an attempt to utilize the DDC formalism to deal with the effects of a thermal bath on the interatomic interaction between two ground-state atoms, we  first generalize the fourth-order DDC formalism  in vacuum~\cite{Zhou21} to in a thermal bath. For simplicity, we start with the model of atoms linearly coupled to a massless scalar field and derive basic formulas for the contribution of thermal fluctuations [tf-contribution] and that of the radiation reaction of atoms [rr-contribution], with which we then calculate the interatomic interaction energy from both contributions and analyze its behaviors in all typical regions in detail. Here let us note that the scalar field model has been widely used in the study of radiative properties of atoms in various circumstances in recent years~\cite{Audretsch95,Audretsch951,Rizzuto16,Zhou21,Iliadakis95,Rizzuto07,Rizzuto091,Zhang11,Rizzuto11,Zhu07,Davies88,Arias16,Menezes18,Zhou20,Zhou16,Zhou17,Zhou1897,Menezes17}.

\section{General formalism of the contributions of thermal fluctuations and radiation reaction of atoms to interatomic interaction}\label{DDC}

We consider that, two spatially separated atoms are interacting with  massless scalar fields in a thermal bath at an arbitrary temperature $T$.
Each atom, labeled by $\xi$~$(\xi={{A}}$ or ${{B}})$, is assumed to be of two energy levels with an energy gap  $\omega_{\xi}$,  and $|g_{\xi}\rangle$ and $|e_{\xi}\rangle$ represent atomic ground state and excited state, with energies $-\frac{\omega_{\xi}}{2}$ and $+\frac{\omega_{\xi}}{2}$ respectively.
The total Hamiltonian describing the ``atoms+field" system is given by
\bea
H(\tau)=H_{{S}}(\tau)+H_{{F}}(\tau)+H_{{I}}(\tau)\;,
\label{total}
\eea
with respect to the proper time of the atoms $\tau$.
Here, $H_{{S}}(\tau)$ and $H_{{F}}(\tau)$ are the free Hamiltonian of the two-atom system and that of the scalar fields, which can be written as
\bea
H_{{S}}(\tau)&=&\omega_{{A}} R^{{A}}_{3}(\tau)+\omega_{{B}} R^{{B}}_{3}(\tau)\;,\label{hs}\\
H_{{F}}(\tau)&=&\int {\mathrm{d}}^3{{\bm{k}}}\;\omega_{\bm{k}}a^{\dag}_{\bm{k}}(t(\tau))a_{\bm{k}}(t(\tau))\frac{{\mathrm{d}}t}{{\mathrm{d}}\tau}\;,\label{hf}
\eea
where $R^{\xi}_{3}=\frac{1}{2}(|e_{\xi}\rangle\langle e_{\xi}|-|g_{\xi}\rangle\langle g_{\xi}|)$,  and $a^{\dag}_{\bm{k}}$ and $a_{\bm{k}}$ are respectively the creation and annihilation operators of the field modes with momentum $\bm{k}$.
The four-dimensional coordinates are denoted by $x=(t,\bm{x})$,
and the atom-field interaction Hamiltonian $H_{{I}}(\tau)$ is expressed as
\bea
H_{{I}}(\tau)=\mu R^{{A}}_{2}(\tau)\phi(x_{{A}}(\tau))+\mu R^{{B}}_{2}(\tau)\phi(x_{{B}}(\tau))\;.
\label{HI}
\eea
Here $\mu$ is a very small coupling constant, 
$R^{\xi}_{2}=\frac{{\mathrm{i}}}{2}(R^{\xi}_{-}-R^{\xi}_{+})$, with $R^{\xi}_{+}=|e_{\xi}\rangle\langle g_{\xi}|$ and $R^{\xi}_{-}=|g_{\xi}\rangle\langle e_{\xi}|$ being atomic raising and lowering operators,
and $\phi(x)$ is the operator of scalar fields,
\bea
\phi(x)=\int {\mathrm{d}}^3{\bm{k}}\; g_{\bm{k}}[a_{\bm{k}}(t)f_{\bm{k}}(\bm{x})+a^{\dag}_{\bm{k}}(t)f^{*}_{\bm{k}}(\bm{x})]\label{phi}\;,
\eea
where $g_{\bm{k}}=[2\omega_{\bm{k}}(2\pi)^3]^{-1/2}$, and $f_{\bm{k}}(\bm{x})$ is the scalar-field mode with momentum $\bm{k}$.
With the Hamiltonians of the ``atoms+field" system in Eqs.~(\ref{total})-(\ref{HI}), we shall first establish the general fourth-order DDC formalism for the contributions of thermal fluctuations and radiation reaction of atoms to the interatomic interaction energy.

Following similar procedures we have adopted in Ref.~\cite{Zhou21} for the interatomic interaction energy between two ground-state atoms in  vacuum, we  obtain and solve the Heisenberg  equations of  motion of the dynamical variables of the atoms and the scalar fields, and separate each solution into a free part and a source part. We then perturbatively expand each source part with respect to the small coupling constant $\mu$.
By exploiting the symmetric ordering between the operators of the atoms and the fields~\cite{DDC82,DDC84}, we evaluate the contributions of the free field and the source field to the variation rate of the Hamiltonian of the two-atom system.
After taking the average of them over the  thermal state of the scalar fields $|\beta\rangle$, we can pick out the following effective Hamiltonian of the thermal fluctuations contribution (tf-contribution) caused by the free field, and that of the radiation reaction contribution (rr-contribution) caused by the source field,
\bea
&&(H_{{S}}(\tau))^{{\mathrm{eff}}}_{{\mathrm{tf}}}\nonumber\\
&=&\frac{1}{8}{\mathrm{i}}\mu^4\int_{\tau_0}^{\tau}{\mathrm{d}}\tau_1\int_{\tau_0}^{\tau_1}{\mathrm{d}}\tau_2\int_{\tau_0}^{\tau_2}{\mathrm{d}}\tau_3
\left\langle\beta\left|\left\{\phi^{{f}}(x_{{A}}(\tau)),\phi^{{f}}(x_{{B}}(\tau_3))\right\}\left[\phi^{{f}}(x_{{B}}(\tau_2)),\phi^{{f}}(x_{{A}}(\tau_1))\right]\right|\beta\right\rangle \nonumber \\
&&~\times \left[R_2^{{{A,f}}}(\tau),R_2^{{{A,f}}}(\tau_1)\right]\left[R_2^{{{B,f}}}(\tau_3),R_2^{{{B,f}}}(\tau_2)\right] \nonumber \\
&&+\text{${{A}}\rightleftharpoons {{B}}$ {term}}\;,
\label{1tf}
\eea
and
{\allowdisplaybreaks[3]
\bea
&&(H_{{S}}(\tau))^{\mathrm{{eff}}}_{\mathrm{{rr}}}\nonumber\\
&=&\frac{1}{8}{\mathrm{i}}\mu^4\int_{\tau_0}^{\tau}{\mathrm{d}}\tau_1\int_{\tau_0}^{\tau_1}{\mathrm{d}}\tau_2\int_{\tau_0}^{\tau_2}{\mathrm{d}}\tau_3
\left\langle\beta\left|\left[\phi^{{f}}(x_{{B}}(\tau_2)),\phi^{{f}}(x_{{A}}(\tau_1))\right]\left[\phi^{{f}}(x_{{B}}(\tau_3)),\phi^{{f}}(x_{{A}}(\tau))\right]\right|\beta\right\rangle \nonumber \\
&&\ \times \left\{R_2^{{{A,f}}}(\tau_1),R_2^{{{A,f}}}(\tau)\right\}\left[R_2^{{{B,f}}}(\tau_2),R_2^{{{B,f}}}(\tau_3)\right] \nonumber \\
&&+\frac{1}{8}{\mathrm{i}}\mu^4\int_{\tau_0}^{\tau}{\mathrm{d}}\tau_1\int_{\tau_0}^{\tau_1}{\mathrm{d}}\tau_2\int_{\tau_0}^{\tau_2}{\mathrm{d}}\tau_3
\left\langle\beta\left|\left[\phi^{{f}}(x_{{A}}(\tau_1)),\phi^{{f}}(x_{{B}}(\tau_3))\right]\left[\phi^{{f}}(x_{{B}}(\tau_2)),\phi^{{f}}(x_{{A}}(\tau))\right]\right|\beta\right\rangle \nonumber \\
&&\ \times \left\{R_2^{{{A,f}}}(\tau_1),R_2^{{{A,f}}}(\tau)\right\}\left[R_2^{{{B,f}}}(\tau_2),R_2^{{{B,f}}}(\tau_3)\right]\nonumber \\
&&+\frac{1}{8}{\mathrm{i}}\mu^4\int_{\tau_0}^{\tau}{\mathrm{d}}\tau_1\int_{\tau_0}^{\tau_1}{\mathrm{d}}\tau_2\int_{\tau_0}^{\tau_2}{\mathrm{d}}\tau_3
\left\langle\beta\left|\left[\phi^{{f}}(x_{{A}}(\tau_3)),\phi^{{f}}(x_{{B}}(\tau_2))\right]\left[\phi^{{f}}(x_{{B}}(\tau_1)),\phi^{{f}}(x_{{A}}(\tau))\right]\right|\beta\right\rangle \nonumber \\
&&\ \times \left\{R_2^{{{A,f}}}(\tau_3),R_2^{{{A,f}}}(\tau)\right\}\left[R_2^{{{B,f}}}(\tau_1),R_2^{{{B,f}}}(\tau_2)\right]\nonumber \\
&&+\frac{1}{8}{\mathrm{i}}\mu^4\int_{\tau_0}^{\tau}{\mathrm{d}}\tau_1\int_{\tau_0}^{\tau}{\mathrm{d}}\tau_2\int_{\tau_0}^{\tau_2}{\mathrm{d}}\tau_3
\left\langle\beta\left|\left[\phi^{{f}}(x_{{B}}(\tau_1)),\phi^{{f}}(x_{{A}}(\tau))\right]\left[\phi^{{f}}(x_{{B}}(\tau_3)),\phi^{{f}}(x_{{A}}(\tau_2))\right]\right|\beta\right\rangle \nonumber \\
&&\ \times \left[R_2^{{{A,f}}}(\tau),R_2^{{{A,f}}}(\tau_2)\right]\left\{R_2^{{{B,f}}}(\tau_1),R_2^{{{B,f}}}(\tau_3)\right\}\nonumber \\
&&+\frac{1}{8}{\mathrm{i}}\mu^4\int_{\tau_0}^{\tau}{\mathrm{d}}\tau_1\int_{\tau_0}^{\tau_1}{\mathrm{d}}\tau_2\int_{\tau_0}^{\tau}{\mathrm{d}}\tau_3
\left\langle\beta\left|\left\{\phi^{{f}}(x_{{B}}(\tau_2)),\phi^{{f}}(x_{{A}}(\tau_3))\right\}\left[\phi^{{f}}(x_{{B}}(\tau_1)),\phi^{{f}}(x_{{A}}(\tau))\right]\right|\beta\right\rangle \nonumber \\
&&\ \times \left[R_2^{{{A,f}}}(\tau),R_2^{{{A,f}}}(\tau_3)\right]\left[R_2^{{{B,f}}}(\tau_2),R_2^{{{B,f}}}(\tau_1)\right]\nonumber \\
&&+\frac{1}{8}{\mathrm{i}}\mu^4\int_{\tau_0}^{\tau}{\mathrm{d}}\tau_1\int_{\tau_0}^{\tau_1}{\mathrm{d}}\tau_2\int_{\tau_0}^{\tau_1}{\mathrm{d}}\tau_3
\left\langle\beta\left|\left[\phi^{{f}}(x_{{B}}(\tau_2)),\phi^{{f}}(x_{{A}}(\tau_1))\right]\left[\phi^{{f}}(x_{{B}}(\tau_3)),\phi^{{f}}(x_{{A}}(\tau))\right]\right|\beta\right\rangle \nonumber \\
&&\ \times \left\{R_2^{{{A,f}}}(\tau_1),R_2^{{{A,f}}}(\tau)\right\}\left[R_2^{{{B,f}}}(\tau_3),R_2^{{{B,f}}}(\tau_2)\right]\nonumber \\
&&+\text{${{A}}\rightleftharpoons {{B}}$ {terms}}\;,\label{1rr}
\eea}where
\bea
\phi^{{f}}(x)=\int {\mathrm{d}}^3{\bm{k}}\ g_{\bm{k}}[a_{\bm{k}}(t_0)\mathrm{e}^{-{\mathrm{i}}\omega_{\bm{k}}(t-t_0)}f_{\bm{k}}(\bm{x})
+a^{\dag}_{\bm{k}}(t_0)\mathrm{e}^{{\mathrm{i}}\omega_{\bm{k}}(t-t_0)}f^{*}_{\bm{k}}(\bm{x})]\;,\label{free-field}
\eea
is the free scalar-field operator, and
\beq
R^{\xi,{{f}}}_{2}(\tau)=\frac{{\mathrm{i}}}{2}\left[R^{\xi}_{-}(\tau_0)\mathrm{e}^{-{\mathrm{i}}\omega_{\xi}(\tau-\tau_0)}-R^{\xi}_{+}(\tau_0)\mathrm{e}^{{\mathrm{i}}\omega_{\xi}(\tau-\tau_0)}\right]\;,\label{R2f}
\eeq
is the free part of atomic operator $R^{\xi}_{2}(\tau)$.
After taking the expectation values of the two effective Hamiltonians, $(H_{{S}}(\tau))^{{\mathrm{eff}}}_{{\mathrm{tf}}}$ and $(H_{{S}}(\tau))^{{\mathrm{eff}}}_{{\mathrm{rr}}}$ in Eqs. (\ref{1tf}) and (\ref{1rr}), over the two-atom state,  $|g_{{A}}g_{{B}}\rangle$, and doing some simplifications, we finally arrive at the general formula for the tf-contribution to the interatomic interaction energy for two ground-state atoms,
\bea
(\delta E)_{\mathrm{{tf}}}&=&\langle g_{{A}}g_{{B}}|(H_{{S}}(\tau))^{{\mathrm{eff}}}_{{\mathrm{tf}}}|g_{{A}}g_{{B}}\rangle\nonumber\\
&=&2{\mathrm{i}}\mu^4\int_{\tau_0}^{\tau}{\mathrm{d}}\tau_1\int_{\tau_0}^{\tau_1}{\mathrm{d}}\tau_2\int_{\tau_0}^{\tau_2}{\mathrm{d}}\tau_3C^{{F}}_{\beta}(x_{{A}}(\tau),x_{{B}}(\tau_3))\chi^{{F}}_{\beta}(x_{{A}}(\tau_1),x_{{B}}(\tau_2))\chi^{{A}}(\tau,\tau_1)\chi^{{B}}(\tau_2,\tau_3)\nonumber\\
&&+\text{${{A}}\rightleftharpoons {{B}}$ {term}}\;,
\label{2tf}
\eea
and that of the rr-contribution,
{\allowdisplaybreaks[3]
\bea
&&(\delta E)_{\mathrm{{rr}}}\nonumber\\
&=&\langle g_{{A}}g_{{B}}|(H_{{S}}(\tau))^{{\mathrm{eff}}}_{\mathrm{{rr}}}|g_{{A}}g_{{B}}\rangle\nonumber\\
&=&2{\mathrm{i}}\mu^4\int_{\tau_0}^{\tau}{\mathrm{d}}\tau_1\int_{\tau_0}^{\tau_1}{\mathrm{d}}\tau_2\int_{\tau_0}^{\tau_2}{\mathrm{d}}\tau_3\chi^{{F}}_{\beta}(x_{{A}}(\tau),x_{{B}}(\tau_3))\chi^{{F}}_{\beta}(x_{{A}}(\tau_1),x_{{B}}(\tau_2))C^{{A}}(\tau,\tau_1)\chi^{{B}}(\tau_2,\tau_3)\nonumber\\
&&+2{\mathrm{i}}\mu^4\int_{\tau_0}^{\tau}{\mathrm{d}}\tau_1\int_{\tau_0}^{\tau_1}{\mathrm{d}}\tau_2\int_{\tau_0}^{\tau_2}{\mathrm{d}}\tau_3\chi^{{F}}_{\beta}(x_{{A}}(\tau_1),x_{{B}}(\tau_3))\chi^{{F}}_{\beta}(x_{{B}}(\tau_2),x_{{A}}(\tau))C^{{A}}(\tau,\tau_1)\chi^{{B}}(\tau_2,\tau_3)\nonumber\\
&&+2{\mathrm{i}}\mu^4\int_{\tau_0}^{\tau}{\mathrm{d}}\tau_1\int_{\tau_0}^{\tau_1}{\mathrm{d}}\tau_2\int_{\tau_0}^{\tau_2}{\mathrm{d}}\tau_3\chi^{{F}}_{\beta}(x_{{A}}(\tau_3),x_{{B}}(\tau_2))\chi^{{F}}_{\beta}(x_{{B}}(\tau_1),x_{{A}}(\tau))C^{{A}}(\tau,\tau_3)\chi^{{B}}(\tau_1,\tau_2)\nonumber\\
&&+2{\mathrm{i}}\mu^4\int_{\tau_0}^{\tau}{\mathrm{d}}\tau_1\int_{\tau_0}^{\tau}{\mathrm{d}}\tau_2\int_{\tau_0}^{\tau_2}{\mathrm{d}}\tau_3\chi^{{F}}_{\beta}(x_{{A}}(\tau_2),x_{{B}}(\tau_3))\chi^{{F}}_{\beta}(x_{{A}}(\tau),x_{{B}}(\tau_1))\chi^{{A}}(\tau,\tau_2)C^{{B}}(\tau_1,\tau_3)\nonumber\\
&&+2{\mathrm{i}}\mu^4\int_{\tau_0}^{\tau}{\mathrm{d}}\tau_1\int_{\tau_0}^{\tau_1}{\mathrm{d}}\tau_2\int_{\tau_0}^{\tau}{\mathrm{d}}\tau_3C^{{F}}_{\beta}(x_{{B}}(\tau_2),x_{{A}}(\tau_3))\chi^{{F}}_{\beta}(x_{{B}}(\tau_1),x_{{A}}(\tau))\chi^{{A}}(\tau_3,\tau)\chi^{{B}}(\tau_1,\tau_2)\nonumber\\
&&+2{\mathrm{i}}\mu^4\int_{\tau_0}^{\tau}{\mathrm{d}}\tau_1\int_{\tau_0}^{\tau_1}{\mathrm{d}}\tau_2\int_{\tau_0}^{\tau_1}{\mathrm{d}}\tau_3\chi^{{F}}_{\beta}(x_{{A}}(\tau),x_{{B}}(\tau_3))\chi^{{F}}_{\beta}(x_{{A}}(\tau_1),x_{{B}}(\tau_2))C^{{A}}(\tau,\tau_1)\chi^{{B}}(\tau_3,\tau_2)\nonumber\\
&&+\text{${{A}}\rightleftharpoons {{B}}$ {terms}}\;,
\label{2rr}
\eea}where $C^{\xi}(\tau,\tau')$ and $\chi^{\xi}(\tau,\tau')$ are the statistical functions of the atoms in the ground state, defined by
\bea
C^{\xi}(\tau,\tau')&\equiv&\frac{1}{2}\langle g_{\xi}|\{R^{\xi,{{f}}}_{2}(\tau),R^{\xi,{{f}}}_{2}(\tau')\}|g_{\xi}\rangle\;,\label{atomC}\\
\chi^{\xi}(\tau,\tau')&\equiv&\frac{1}{2}\langle g_{\xi}|[R^{\xi,{{f}}}_{2}(\tau),R^{\xi,{{f}}}_{2}(\tau')]|g_{\xi}\rangle\label{atomChi}\;,
\eea
and
$C^{{F}}_{\beta}(x_{{A}}(\tau),x_{{B}}(\tau'))$ and $\chi^{{F}}_{\beta}(x_{{A}}(\tau),x_{{B}}(\tau'))$ are the correlation functions of the scalar fields in the thermal state $|\beta\rangle$, defined by
\bea
C^{{F}}_{\beta}(x_{{A}}(\tau),x_{{B}}(\tau'))&\equiv&\frac{1}{2}\langle \beta|\{\phi^{{{f}}}(x_{{A}}(\tau)),\phi^{{{f}}}(x_{{B}}(\tau'))\}|\beta\rangle\;,\label{fieldC}\\
\chi^{{F}}_{\beta}(x_{{A}}(\tau),x_{{B}}(\tau'))&\equiv&\frac{1}{2}\langle\beta|[\phi^{{{f}}}(x_{{A}}(\tau)),\phi^{{{f}}}(x_{{B}}(\tau'))]|\beta\rangle\;.\label{fieldChi}
\eea
Note here that, in order to obtain Eqs.~(\ref{2tf}) and (\ref{2rr}), we have simplified every scalar-field four-point correlation function of the effective Hamiltonians in Eqs.~(\ref{1tf}) and (\ref{1rr}), into products of two two-point correlation functions, and the details on the simplifications are given in Appendix.~\ref{two-four-point-correlation-field}.


\section{Contributions of thermal fluctuations and radiation reaction to the interaction of two static ground-state atoms} \label{gg-interaction}

In this section, we shall use the general formalism established in the proceeding  section to  calculate the interatomic interaction energy between two static ground-state atoms in a thermal bath.
The trajectories of the two atoms at rest are depicted by
\bea
x_{{A}}=(\tau,0,0,0),\quad x_{{B}}=(\tau,0,0,L),
\eea
where $L~(>0$) denotes the interatomic separation,
and along these trajectories of the atoms, the correlation functions of the scalar fields defined in Eqs.~(\ref{fieldC}) and (\ref{fieldChi}) are respectively
\bea
C^{{F}}_{\beta}(x_{{A}}(\tau),x_{{B}}(\tau'))&=&\frac{1}{4\pi^2L}\int_{0}^{\infty}{\mathrm{d}}\omega\left(1+\frac{2}{{\mathrm{e}}^{\omega/T}-1}\right)\sin(\omega L)\cos{[\omega(\tau-\tau')]}\;,\label{Cth}\\
\chi^{{F}}_{\beta}(x_{{A}}(\tau),x_{{B}}(\tau'))&=&-\frac{{\mathrm{i}}}{4\pi^2L}\int_{0}^{\infty}{\mathrm{d}}\omega~ \sin(\omega L)\sin{[\omega(\tau-\tau')]}\;.\label{chith}
\eea
For atomic statistical functions defined in Eqs.~(\ref{atomC}) and (\ref{atomChi}), we  obtain, with Eq.~(\ref{R2f}),
\bea
C^{\xi}(\tau,\tau')&=&\frac{1}{4}\cos{[\omega_{\xi}(\tau-\tau')]}\;,\label{atomC1}\\
\chi^{\xi}(\tau,\tau')&=&-\frac{{\mathrm{i}}}{4}\sin{[\omega_{\xi}(\tau-\tau')]}\;,\label{atomChi1}
\eea
which are valid when the two-level atom is in its ground state $|g_{\xi}\rangle$.
Substituting Eqs.~(\ref{Cth})-(\ref{atomChi1}) into Eqs. (\ref{2tf}) and (\ref{2rr}), and performing the triple integrations  with respect to $\tau_1$, $\tau_2$ and  $\tau_3$, we have respectively, for an  infinitely long time interval $\tau-\tau_0$, the tf-contribution and the rr-contribution
\bea
(\delta E)_{{\mathrm{tf}}}=-\frac{\mu^4}{64\pi^4L^2}\int_{0}^{\infty}{\mathrm{d}}\omega_1\int_{0}^{\infty}{\mathrm{d}}\omega_2\frac{\omega_{{A}} \omega_{{B}}\omega_2\sin(\omega_1L)\sin(\omega_2 L)}{(\omega_1^2-\omega_{{A}}^2)(\omega_1^2-\omega_{{B}}^2)(\omega_2^2-\omega_1^2)}\left(1+\frac{2}{{\mathrm{e}}^{\omega_1/T}-1}\right)\;,
\label{tf3}
\eea
and
\bea
(\delta E)_{{\mathrm{rr}}}&=&(\delta E)_{\mathrm{{tf}}}-\frac{\mu^4}{32\pi^4L^2}\int_{0}^{\infty}{\mathrm{d}}\omega_1\int_{0}^{\infty}{\mathrm{d}}\omega_2\frac{\omega_1\omega_2(\omega_1^2-\omega_{{A}}^2-\omega_{{A}}\omega_{{B}}-\omega_{{B}}^2)
}{(\omega_{{A}}+\omega_{{B}})(\omega_1^2-\omega_{{A}}^2)(\omega_1^2-\omega_{{B}}^2)(\omega_2^2-\omega_1^2)}\nonumber\\
&&\;\;\qquad\qquad\qquad\qquad\qquad\qquad\qquad \times\sin(\omega_1L)\sin(\omega_2L)\;.
\label{rr3}
\eea
The above results can be further simplified by performing the $\omega_2-$integration with the residue theorem to yield the final expressions of the tf-contribution and the rr-contribution
\bea
(\delta E)_{{\mathrm{tf}}}&=&-\frac{\mu^4}{512\pi^2L^2}\frac{\omega_{{B}}\cos{(2\omega_{{A}} L)}-\omega_{{A}}\cos{(2\omega_{{B}} L)}}{\omega_{{A}}^2-\omega_{{B}}^2}
-\frac{\mu^4}{256\pi^3L^2}\int_{0}^{\infty}{\mathrm{d}}u\alpha_{{A}}(\mathrm{i}u)\alpha_{{B}}(\mathrm{i}u){\mathrm{e}}^{-2uL}\nonumber\\
&&-\frac{\mu^4}{128\pi^3L^2}\int_{0}^{\infty}{\mathrm{d}}u\alpha_{{A}}(u)\alpha_{{B}}(u)\frac{\sin{(2u L)}}{{\mathrm{e}}^{u/T}-1}
\label{tf2}
\eea
and
\bea
(\delta E)_{\mathrm{{rr}}}&=&\frac{\mu^4}{512\pi^2L^2}\frac{\omega_{{B}}\cos{(2\omega_{{A}} L)}-\omega_{{A}}\cos{(2\omega_{{B}} L)}}{\omega_{{A}}^2-\omega_{{B}}^2}
-\frac{\mu^4}{256\pi^3L^2}\int_{0}^{\infty}{\mathrm{d}}u\alpha_{{A}}(\mathrm{i}u)\alpha_{{B}}(\mathrm{i}u){\mathrm{e}}^{-2uL}\nonumber\\
&&-\frac{\mu^4}{128\pi^3L^2}\int_{0}^{\infty}{\mathrm{d}}u\alpha_{{A}}(u)\alpha_{{B}}(u)\frac{\sin{(2u L)}}{{\mathrm{e}}^{u/T}-1}
\label{rr2}
\eea
with $\alpha_{\xi}(\mathrm{i}u)=\omega_{\xi}/(\omega_{\xi}^2+u^2)$,
{both of which  are composed of three terms with the first two temperature-independent  terms  reflecting effects of zero-point fluctuations and the third  temperature-dependent  term representing thermal revisions. It is also worth noting that every term of Eqs.~(\ref{tf2}) and (\ref{rr2}) has a prefactor $\sim L^{-2}$, and it arises from the prefactor $L^{-1}$ in the field correlation functions (Eqs.~(\ref{Cth}) and (\ref{chith})) as the tf- and rr-contribution depend on the product of two field correlation functions (see Eqs.~(\ref{2tf}) and (\ref{2rr})). Generally, the tf- and rr-contribution are distinctive due to  that the first  term in  Eq.~(\ref{tf2}) and Eq.~(\ref{rr2}) which is oscillatory is of the same magnitude but an opposite sign. When added up, the first terms in these two equations cancel out perfectly, and it means that this part of the effects of zero-point fluctuations in the tf- and rr-contribution is cancelled. As a result, the total interaction energy follows,}
\bea
(\delta E)_{\mathrm{tot}}&=&-\frac{\mu^4}{128\pi^3L^2}\int_{0}^{\infty}{\mathrm{d}}u\frac{\omega_{{A}}\omega_{{{B}}}\ }{(\omega_{{{A}}}^2+u^2)(\omega_{{{B}}}^2+u^2)}{\mathrm{e}}^{-2uL}\nonumber\\
&&-\frac{\mu^4}{64\pi^3L^2}\int_{0}^{\infty}{\mathrm{d}}u\frac{\omega_{{{A}}}\omega_{{{B}}}\sin{(2u L)}}{(\omega_{{{A}}}^2-u^2)(\omega_{{{B}}}^2-u^2)({\mathrm{e}}^{u/T}-1)}\;,\quad\label{tb}
\eea
{which contains two terms with the first temperature-independent term being exactly the interaction energy of two static ground-state atoms in vacuum and the other  temperature-dependent term giving the thermal corrections. }

It is difficult to analytically evaluate the integrals in Eqs.~(\ref{tf2})-(\ref{tb}). Fortunately, however, analytical approximate results are obtainable in some special cases.
For simplicity, we assume in the following discussions that the two atoms are identical with the transition frequency denoted by $\omega_0$, and divide our analysis into  the van der Waals regime where the interatomic separation $L$ is much smaller than the atomic transition wavelength $\lambda$, defined by $\lambda=2\pi\omega_0^{-1}$, and the Casimir-Polder regime  where $L$  is much larger than the atomic transition wavelength $\lambda$. {As indicated in the general expressions of the tf- and rr-contribution Eqs.~(\ref{tf2}) and (\ref{rr2}), for the interaction energy of two atoms in a thermal bath, besides the two length scales $L$ and $\lambda$, another length scale, i.e., the characteristic wavelength of photons in the thermal bath $\beta$ singles out. As a result, the van der Waals region $L\ll\lambda$ and the Casimir-Polder region $L\gg\lambda$ are respectively further divided into three typical regions, and the tf- and rr-contribution (Eqs.~(\ref{tf2}) and (\ref{rr2})), which can be  expressed in terms of some complicated special functions after doing the integrations,  can be further approximated in polynomials of $L$. We next discuss those results one by one.  }

\subsection{The van der Waals interaction in a thermal bath}
This regime can be further divided into three sub-regimes depending on the relationship among three characteristic length scales, i.e.,  the interatomic separation $L$,
the thermal wavelength of the thermal bath, $\beta=T^{-1}$, and atomic transition wavelength $\lambda$. We now start with $L\ll\lambda\ll\beta$.

\subsubsection{The van der Waals interaction in the region $L\ll\lambda\ll\beta$}

When $L\ll\lambda\ll\beta$, the tf-contribution and the rr-contribution to the interatomic van der Waals interaction, according to Eqs. (\ref{tf2}) and (\ref{rr2}), are respectively approximated by
\bea
(\delta E)_{{\mathrm{tf}}}&\simeq&\frac{\mu^4}{256\pi^3L}-\frac{\mu^4 T^2}{384\pi\omega_0^2 L}-\frac{\mu^4\omega_0^2L}{192\pi^3}\ln{(\omega_0L)}\;,\label{short1tf}\\
(\delta E)_{{\mathrm{rr}}}&\simeq&-\frac{\mu^4}{512\pi^2\omega_0 L^2}+\frac{\mu^4}{256\pi^3L}-\frac{\mu^4 T^2}{384\pi \omega_0^2 L}-\frac{\mu^4\omega_0^2L}{192\pi^3}\ln{(\omega_0L)}\;.\label{short1rr}
\eea
In each of these two equations, the first term on the right plays a dominant role, and the thermal corrections dependent on the temperature appear in higher orders, which are negligible.
Taking the derivative of the above leading terms with respect to the interatomic separation, we find that the thermal fluctuations lead to a repulsive van der Waals force between two atoms, which is proportional to $L^{-2}$, while the radiation reaction of atoms yields an attractive force which is proportional to $L^{-3}$.
{Let us note that  different  scaling of $L$ in these two leading terms is caused by the difference in the sign of oscillatory terms in Eqs. (\ref{tf2}) and (\ref{rr2}). To be specific, the oscillatory term in the tf-contribution (Eq. (\ref{tf2})) cancels the other temperature-independent term, while that in the rr-contribution (Eq. (\ref{rr2})) coincides it and they jointly lead to the first term in Eq.~(\ref{short1rr}).}
It is easy to see that  $|\delta E|_{{\mathrm{tf}}}\ll|\delta E|_{{\mathrm{rr}}}$,
which means that the rr-contribution dominates over the tf-contribution in this case.
Adding up Eqs. (\ref{short1tf}) and (\ref{short1rr}),  we get
\bea
(\delta E)_{{\mathrm{tot}}}\simeq-\frac{\mu^4}{512\pi^2\omega_0 L^2}+\frac{\mu^4}{128\pi^3L}-\frac{\mu^4 T^2}{192\pi\omega_0^2 L}-\frac{\mu^4\omega_0^2L}{96\pi^3}\ln{(\omega_0L)}\;,\label{short1}
\eea
for the  total van der Waals interaction energy of two ground-state atoms in a thermal bath, which is attractive.

\subsubsection{The van der Waals interaction in the region $L\ll \beta\ll \lambda$}

When $L\ll \beta\ll \lambda$, we simplify the tf-contribution and the rr-contribution in Eqs.~(\ref{tf2}) and (\ref{rr2}) into the following two expressions,
\bea
(\delta E)_{{\mathrm{tf}}}&\simeq&-\frac{\mu^4}{256\pi^3L}-\frac{\mu^4\omega_0^2}{384\pi^3 T^2 L}\ln{\left(\frac{\omega_0}{T}\right)}\;,\label{short2tf}\\
(\delta E)_{{\mathrm{rr}}}&\simeq&-\frac{\mu^4}{512\pi^2\omega_0 L^2}-\frac{\mu^4}{256\pi^3L}-\frac{\mu^4\omega_0^2}{384\pi^3 T^2 L}\ln{\left(\frac{\omega_0}{T}\right)}\;.\label{short2rr}
\eea
Comparing the above results 
with their counterparts in the case of $L\ll\lambda\ll \beta$, i.e. Eqs. (\ref{short1tf}) and (\ref{short1rr}),
we find that the first (leading) term of the tf-contribution in Eq.~(\ref{short2tf}), as well as the second (subleading) term of the rr-contribution in Eq.~(\ref{short2rr}), which are independent of the temperature, remarkably change sign.
{This sign-changing phenomenon results from that the sole temperature-dependent term in Eqs.~(\ref{tf2}) and (\ref{rr2}), in the limit of $L\ll\beta\ll \lambda$ also generates the temperature-independent behaviors of $L^{-1}$, and thus suggests that, in sharp contrast to the low-temperature case of $\lambda\ll\beta$, the thermal radiation can also induce the temperature-independent corrections to the van der Waals interaction in the high-temperature case, i.e., when the thermal wavelength $\beta$ is much smaller than the transition wavelength of atoms $\lambda$.}
Consequently, the tf-contribution in the $L\ll \beta\ll \lambda$ region leads to an attractive van der Waals force, as apposed to  a repulsive force in the region $L\ll\lambda\ll\beta$, and the rr-contribution leads to a slight enhancement  for the van der Waals interaction energy, rather than a slight reduction in the region $L\ll\lambda\ll\beta$.

The summation of the above two equations, Eqs.~(\ref{short2tf}) and (\ref{short2rr}), gives rise to the total van der Waals interaction energy in the region $L\ll \beta\ll \lambda$,
\bea
(\delta E)_{{\mathrm{tot}}}\simeq-\frac{\mu^4}{512\pi^2\omega_0 L^2}-\frac{\mu^4}{128\pi^3L}-\frac{\mu^4\omega_0^2}{192\pi^3 T^2 L}\ln{\left(\frac{\omega_0}{T}\right)}\;.
\eea
Obviously, though the dominant behavior of the tf-contribution is altered by the temperature,
the leading term of the total interaction energy  comes from the rr-contribution in Eq. (\ref{short2rr}), which however is unaffected by the temperature.
As a result, the total interatomic van der Waals force in this case is, to the leading order, still an attractive force proportional to $L^{-3}$.
{In the present region and the region of $L\ll\lambda\ll\beta$, the contribution of radiation reaction of atoms to the interaction energy is always larger than that of field fluctuations, and this is physically understandable  as the  propagation distance of the  photons of  the radiative field emitted from one atom to the other atom is very  short when two atoms are very close. }

\subsubsection{The van der Waals interaction in the region $\beta\ll L\ll \lambda$}

The third region for the van der Waals interaction we are to examine is  $\beta\ll L\ll \lambda$, within which the tf-contribution and the rr-contribution in Eqs. (\ref{tf2}) and (\ref{rr2}) are approximated as
\bea
(\delta E)_{\mathrm{{tf}}}&\simeq&-\frac{\mu^4}{256\pi^3L}-\frac{\mu^4\omega_0^2L^2T}{384\pi^2}\;,\label{intermediate2tf}\\
(\delta E)_{\mathrm{{rr}}}&\simeq&-\frac{\mu^4}{512\pi^2\omega_0 L^2}-\frac{\mu^4}{256\pi^3L}-\frac{\mu^4\omega_0^2L^2T}{384\pi^2}\;,
\label{intermediate2rr}
\eea
and they jointly result in the following total van der Waals interaction energy,
\bea
(\delta E)_{\mathrm{{tot}}}\simeq-\frac{\mu^4}{512\pi^2\omega_0 L^2}-\frac{\mu^4}{128\pi^3L}-\frac{\mu^4\omega_0^2L^2T}{192\pi^2}\;.\label{intermediate2}
\eea
In order to elucidate   the behaviors of the tf-contribution and the rr-contribution to the interatomic van der Waals interaction energy in the $\beta\ll L\ll \lambda$  region,
we further divide this region into the following three different subregions, i.e. the subregion of $\beta\ll L\ll\sqrt[3]{\lambda^2\beta}$, $\sqrt[3]{\lambda^2\beta}\ll L\ll\sqrt[4]{\lambda^3\beta}$, and $\sqrt[4]{\lambda^3\beta}\ll L\ll\lambda$.

In the first subregion $\beta\ll L\ll\sqrt[3]{\lambda^2\beta}$, we obtain
\bea
(\delta E)_{{\mathrm{tf}}}&\simeq&-\frac{\mu^4}{256\pi^3L}\;,\\
(\delta E)_{\mathrm{{rr}}}&\simeq&-\frac{\mu^4}{512\pi^2\omega_0 L^2}\;,
\eea
which are exactly the same in the leading order as those in the region $L\ll\beta\ll\lambda$ [see Eqs.~(\ref{short2tf}) and (\ref{short2rr})].
And the tf-contribution and the rr-contribution, in the second subregion $\sqrt[3]{\lambda^2\beta}\ll L\ll\sqrt[4]{\lambda^3\beta}$, can be expressed as
\bea
(\delta E)_{{\mathrm{tf}}}&\simeq&-\frac{\mu^4\omega_0^2L^2T}{384\pi^2}\;, \\
(\delta E)_{{\mathrm{rr}}}&\simeq&-\frac{\mu^4}{512\pi^2\omega_0 L^2}\;,
\eea
respectively.  Here the van der Waals force attributed to the tf-contribution is a repulsive force, with a new behavior of $TL$.
It is easy to verify that $|(\delta E)_{{\mathrm{rr}}}|\gg|(\delta E)_{{\mathrm{tf}}}|$ in the first two subregions, i.e., the rr-contribution weighs much more  than the tf-contribution.
The third subregion comes with $\sqrt[4]{\lambda^3\beta}\ll L\ll\lambda$, and now the tf-contribution plays an equally important role as the rr-contribution does, i.e.
\beq
(\delta E)_{{\mathrm{tf}}}\simeq(\delta E)_{{\mathrm{rr}}}\simeq-\frac{\mu^4\omega_0^2L^2T}{384\pi^2}\;.
\eeq
It is interesting to note that  the interatomic van der Waals energy leads to an attractive  force  in the first two subregions which scales as  $L^{-3}$, and remarkably  a repulsive force in the third subregion, i.e., the $\sqrt[4]{\lambda^3\beta}\ll L\ll\lambda$ region, which scales as $TL$. The crossover of the van der Waals force from attractive to repulsive when the interatomic separation enters the third subregion is a result of the fact that both the tf- and rr-contribution are dominated by the temperature-dependent terms which weigh equally, whereas, in the first two regions $L\ll\lambda\ll\beta$, $L\ll\beta\ll\lambda$, or the first two subregions of $\beta\ll L\ll\lambda$, although the tf-contributions may be dominated by temperature-dependent terms, the rr-contribution, which is often dominated by temperature-independent terms, is much greater than the tf-contribution, and consequently the total interaction energy is still dominated by temperature-independent terms, resulting in an overall attractive force.


{ Some comments are in order on whether or not this repulsive van der Waals force is experimentally detectable. For the remarkable crossover of the van der Waals force from attractive to repulsive to occur, $\beta\ll\lambda$ has to be satisfied, i.e., the thermal-bath temperature $T$ has to be much higher than the critical temperature $T_{{c}}\equiv \hbar\omega_0/k_{\mathrm{B}}$ [in the International System of Units]. For natural atoms with transition frequencies $\omega_0\in[10^{14},10^{17}]s^{-1}$,  $T_{{c}}\sim764K$ at least, which is much higher than the room temperature $T\sim300K$. Thus this repulsive van der Waals force of two natural atoms does not show up near the room temperature. One may use the atoms with relatively lower transition frequencies to reduce the temperature required for a repulsive van der Waals force to happen. However, a smaller $\omega_0$ and a smaller $T$ means a larger interatomic separation since $L\gg\sqrt[4]{\lambda^3\beta}$ is required. For instance, it can be estimated that at the room temperature, if a repulsive van der Waals force is to be generated between two atoms with a transition frequency $\omega_0\sim10^{12}s^{-1}$, the interatomic separation has to be approximately hundreds of micrometers, which makes the force too tiny to detect at present. So, an  experimental detection of the repulsive  van der Waals force relies on  technological advances in precision of  measurement in the future.}


\subsection{The Casimir-Polder interaction in a thermal bath}

In this subsection, we turn  to explore how the interatomic Casimir-Polder interaction energy is affected by the presence of a thermal bath. 
Similar to the discussions on the van der Waals interaction energy above, our analytical results for the interatomic Casimir-Polder interaction energy are approximately obtained in the regions of $\lambda\ll L\ll\beta$, $\lambda\ll\beta\ll L$, and $\beta\ll\lambda\ll L$.

\subsubsection{The Casimir-Polder interaction in the region $\lambda\ll L\ll\beta$}

In the region $\lambda\ll L\ll\beta$, we approximate the tf-contribution and the rr-contribution to the interatomic Casimir-Polder interaction energy, Eqs.~(\ref{tf2}) and (\ref{rr2}), as
\bea
(\delta E)_{\mathrm{{tf}}}&\simeq&\frac{\mu^4}{512\pi^2L}\sin{(2\omega_0 L+\theta_1)}-\frac{\mu^4}{512\pi^3\omega_0^2L^3}-\frac{\mu^4 T^2}{384\pi \omega_0^2 L}\;,\label{intermediate1tf}\\
(\delta E)_{{\mathrm{rr}}}&\simeq&-\frac{\mu^4}{512\pi^2L}\sin{(2\omega_0 L+\theta_1)}-\frac{\mu^4}{512\pi^3\omega_0^2L^3}-\frac{\mu^4 T^2}{384\pi \omega_0^2 L}\;,\label{intermediate1rr}
\eea
with $\theta_1=\arcsin{(1+4\omega_0^2L^2)^{-1/2}}$.
{In sharp contrast to those results in the three van der Waals regions previously discussed, here the first term of Eq.~(\ref{intermediate1tf}) and that of Eq.~(\ref{intermediate1rr}), which are temperature-independent, oscillate with the interatomic separation and they respectively correspond to the first term of the general expression of the tf-contribution Eq.~(\ref{tf2}) and that of the rr-contribution Eq.~(\ref{rr2}) in the limit of $\omega_{\xi}\rightarrow\omega_0$ and $L\gg\lambda$.} 
{The second term 
is also temperature-independent, and it can be derived by replacing $\alpha_{\xi}(\mathrm{i}u)$ in the second term of Eqs.~(\ref{tf2}) and (\ref{rr2}) with $\alpha_{\xi}(0)$ which is reasonable as the exponential $\mathrm{e}^{-2uL}$ in the integrand and the condition $\lambda\ll L$ in the present case ensure the rapid decrease of the integrand with the increase of $u$. The oscillatory term together with the second term in the above results are manifestations of effects of zero-point fluctuations.} 
{The third terms which are proportional to $T^2L^{-1}$, are thermal revisions and they result from the third terms of Eqs.~(\ref{tf2}) and (\ref{rr2}). Comparing the amplitude of the oscillatory term in each of the above two equations with non-oscillatory terms, we find that the former is much greater, indicating that the tf- and rr-contribution to the interaction energy in the region $\lambda\ll L\ll\beta$ can be either positive or negative and even be null,
signaling an either attractive or repulsive and even vanishing force}.

However, this oscillatory separation-dependent behavior does not remain in the total Casimir-Polder interaction energy, as the signs of the oscillatory terms in $(\delta E)_{\mathrm{{tf}}}$ and $(\delta E)_{{\mathrm{rr}}}$ are opposite.
As a result, we obtain the following monotonic behavior of the interatomic Casimir-Polder interaction energy in the region $\lambda\ll L\ll\beta$,
\bea
(\delta E)_{\mathrm{{tot}}}\simeq-\frac{\mu^4}{256\pi^3\omega_0^2L^3}-\frac{\mu^4 T^2}{192\pi \omega_0^2 L}\;,\label{intermediate1}
\eea
which are equally contributed by the tf-contribution and the rr-contribution, Eqs. (\ref{intermediate1tf}) and (\ref{intermediate1rr}).
{Here, the change from the van der Waals behavior of $L^{-2}$ in Eq. (\ref{short1}) to the Casimir-Polder behavior of $L^{-3}$ in Eq. (\ref{intermediate1}) shows the retardation effect as a result of finite light speed.}

\subsubsection{The Casimir-Polder interaction in the region $\lambda\ll\beta\ll L$}

When $\lambda\ll \beta\ll L$, we obtain, with Eqs.~(\ref{tf2}) and (\ref{rr2}), the tf-contribution and rr-contribution to the Casimir-Polder interaction energy,
\bea
(\delta E)_{\mathrm{{tf}}}&\simeq&\frac{\mu^4}{512\pi^2L}\sin{(2\omega_0 L+\theta_1)}-\frac{\mu^4 T}{256\pi^2\omega_0^2L^2}\;,\label{long1tf}\\
(\delta E)_{\mathrm{{rr}}}&\simeq&-\frac{\mu^4}{512\pi^2L}\sin{(2\omega_0 L+\theta_1)}-\frac{\mu^4 T}{256\pi^2\omega_0^2L^2}\;.\label{long1rr}
\eea
We find that, the same oscillatory terms as those in the region $\lambda\ll L\ll\beta$, 
 also appear in the present case, and their amplitudes are {also} much larger than the monotonic temperature-dependent terms.
The summation of the above tf-contribution and rr-contribution,
as a result of the perfect cancellation of the oscillatory terms,
 yields
\bea
(\delta E)_{{\mathrm{tot}}}\simeq-\frac{\mu^4 T}{128\pi^2\omega_0^2L^2}\;,\label{long1}
\eea
which gives the total Casimir-Polder interaction energy of two ground-state atoms  in  the $\lambda\ll \beta\ll L$ region.
{Obviously, this interaction energy equally comes from the tf- and rr-contribution, and it can be easily derived by replacing $\alpha_{\xi}(u)$ with $\alpha_{\xi}(0)$ in the integrand of Eqs.~(\ref{tf2}) and (\ref{rr2}) since an exponential factor $({\mathrm{e}}^{u/T}-1)^{-1}$ is included in the integrand and meanwhile $\omega_0\gg T$ for the present case. As compared with the temperature-independent behavior of $L^{-3}$ in the vacuum case or the thermal case with $\lambda\ll L\ll\beta$, this Casimir-Polder interaction energy is dominated by thermal corrections and it displays the same separation-dependence, $L^{-2}$, just as the van der Waals interaction energy in the vacuum case, suggesting that temperature effects in this case suppress the retardation effect.}


\subsubsection{The Casimir-Polder interaction in the region $\beta\ll\lambda\ll L$}

Finally,  in the $\beta\ll\lambda\ll L$ region,
the tf-contribution and the rr-contribution to the interatomic interaction energy are respectively
\bea
(\delta E)_{{\mathrm{tf}}}\simeq (\delta E)_{{\mathrm{rr}}}\simeq \frac{\mu^4T}{256\pi^2\omega_0L}\sin{(2\omega_0 L+\theta_2)}-\frac{\mu^4 T}{256\pi^2\omega_0^2L^2}
\;,\label{long4tf}
\eea
with $\theta_2=\arcsin{(1+\omega_0^2L^2)^{-1/2}}$.
A remarkable difference, as compared to the case for $\lambda\ll \beta$ [refer to Eqs. (\ref{long1tf}) and (\ref{long1rr})], lies in that the leading oscillatory terms of the tf-contribution and the rr-contribution 
are temperature-dependent and they do not cancel out with each other, leading to the total interatomic Casimir-Polder interaction which is oscillatory,
\bea
(\delta E)_{\mathrm{{tot}}}\simeq\frac{\mu^4T}{128\pi^2\omega_0L}\sin{(2\omega_0 L+\theta_2)}-\frac{\mu^4 T}{128\pi^2\omega_0^2L^2}
\;.\label{long4}
\eea
So, the  total  interatomic Casimir-Polder interaction  oscillates with a decreasing amplitude which is proportional to $TL^{-1}$, and as a result,
the Casimir-Polder force can be either attractive or repulsive, and even vanish, depending on the concrete value of the separation.
For some special interatomic separations, in which the leading oscillatory term vanishes,
the Casimir-Polder interaction energy is proportional to $TL^{-2}$ and corresponds to an attractive force, as  in the  $\lambda\ll\beta\ll L$ region [see  Eq. (\ref{long1})].

{So far, we have derived the interaction energy of two ground-state atoms in a thermal bath and analysed in detail the approximate analytical results in three van der Waals regions and three Casimir-Polder regions. It is worth pointing out that for two atoms with an arbitrary separation in a thermal bath at an arbitrary temperature, though the interaction energy Eqs.~(\ref{tf2})-(\ref{tb}) is generally difficult to simplify, it can be numerically computed after the removing of the removable singularity $u=\omega_{\xi}$. We have done such numerical computations, compared the numerical results with all the approximate analytical results in six typical regions we have reported in the present work, and found that they agree very well.}

\section{Summary}\label{conclusions}

In this paper, we first make a generalization of the fourth-order DDC formalism from vacuum to a thermal bath, and with this general formalism, we then study the interatomic interaction energy between two static ground-state atoms coupled to massless scalar fields at finite temperature $T$.
We analytically calculate and analyze, in six regions depending on the relative magnitudes among  three characteristic physical parameters of the system, i.e. the interatomic separation $L$, the atomic transition wavelength $\lambda$, and the thermal wavelength $\beta$, the separate contribution of thermal fluctuations (tf-contribution) and that of the radiation reaction of atoms (rr-contribution) to the interatomic van der Waals and Casimir-Polder interaction energies, and some important results are summed up as follows.

The analytical results for the van der Waals interaction energy are approximately obtained in the following three regions: $L\ll\lambda\ll\beta$, $L\ll \beta\ll \lambda$ and $\beta\ll L\ll \lambda$, and the third region $\beta\ll L\ll \lambda$ is further divided into  three subregions: $\beta\ll L\ll\sqrt[3]{\lambda^2\beta}$, $\sqrt[3]{\lambda^2\beta}\ll L\ll\sqrt[4]{\lambda^3\beta}$, and $\sqrt[4]{\lambda^3\beta}\ll L\ll\lambda$. We find that, in the first two regions as well as the first two subregions of the third region,
the thermal corrections to the van der Waals interaction are of a high-order effect, and the interatomic interaction in the leading order is the same as that in vacuum, where the rr-contribution plays a dominant role and leads to an attractive force with the behavior of $L^{-3}$.
Remarkably, 
a crossover of the van der Waals force from attractive to repulsive occurs, when the temperature of the thermal bath is sufficiently high such that $\sqrt[4]{\lambda^3\beta}\ll L\ll\lambda$. In this subregion within the $\beta\ll L\ll \lambda$ region, the tf-contribution and the rr-contribution play an equally important role, and  they jointly result in a repulsive van der Waals force which scales as $TL$.

Similarly, analytical results for the Casimir-Polder interaction energy are obtained in regions $\lambda\ll L\ll\beta$, $\lambda\ll \beta\ll L$ and $\beta\ll\lambda\ll L$, and the tf-contribution and the rr-contribution in every region are found to be oscillatory and are equally important. However, these oscillatory separation-dependent behaviors do not carry onto the total interaction energy in the first two regions because of a perfect cancellation of the two oscillatory contributions,  resulting in an attractive interatomic interaction force with a monotonic separation-dependence. Concretely, the force scales as $L^{-4}$  in the first region, which is the same as its counterpart in vacuum, and  scales as $TL^{-3}$ in the second region, which is dominated by the thermal corrections. In the third region, the total interaction energy oscillates with a decreasing amplitude proportional to $TL^{-1}$ and gives rise to a force which can be either attractive or repulsive and even vanishing, depending on the concrete value of the interatomic separation.

\begin{acknowledgments}
This work was supported in part by the NSFC under Grants No. 11690034, No. 12075084,
No. 11875172 and No. 12047551,  and No. 12105061, and the K. C. Wong Magna Fund in Ningbo University.
\end{acknowledgments}

\appendix

\section{Details on the simplifications of the four-point correlation functions of the scalar fields in Eqs.~(\ref{1tf}) and (\ref{1rr}).}\label{two-four-point-correlation-field}

As we have mentioned below Eq.~(\ref{fieldChi}), in order to obtain Eqs.~(\ref{2tf}) and (\ref{2rr}), we have simplified the four-point correlation functions of the free fields in the effective Hamiltonians of Eqs.~(\ref{1tf}) and (\ref{1rr}), into the products of two two-point correlation functions of the scalar fields. We now give the details on the simplifications, i.e., how $\langle \beta|\phi^{{f}}(x_1)\phi^{{f}}(x_2)\phi^{{f}}(x_3)\phi^{{f}}(x_4)|\beta\rangle$ can be expanded in terms of the product of two two-point correlation functions.

The four-point correlation function of the scalar fields in a thermal state, $\langle\beta|\phi^{{f}}(x_1)\phi^{{f}}(x_2)\phi^{{f}}(x_3)\phi^{{f}}(x_4)|\beta\rangle$, is generally given by
\bea
&&\langle\beta|\phi^{{f}}(x_1)\phi^{{f}}(x_2)\phi^{{f}}(x_3)\phi^{{f}}(x_4)|\beta\rangle\nonumber\\
&=&\frac{1}{Z}\sum_{\bm{k}_1,\bm{k}_2,\bm{k}_3,\bm{k}_4}\mathrm{Tr}\biggl\{\mathrm{e}^{-\beta H_{{F}}}
\left[a^{{f}}_{\bm{k}_1}{{f}}_{\bm{k}_1}(x_1)+\mathrm{H. c.}\right]\left[a^{{f}}_{\bm{k}_2}{{f}}_{\bm{k}_2}(x_2)+\mathrm{H. c.}\right]\nonumber\\
&&\quad\qquad\qquad\qquad\qquad \times\left[a^{{f}}_{\mathbf{k}_3}{{f}}_{\mathbf{k}_3}(x_3)+\mathrm{H. c.}\right]\left[a^{{f}}_{\bm{k}_4}{{f}}_{\bm{k}_4}(x_4)+\mathrm{H. c.}\right]\biggl\}\;,
\label{fourpoint}
\eea
where $H_{{F}}=\sum\limits_{\bm{k}}\omega_{\bm{k}}a^{{{f}}\dag}_{\bm{k}}a^{{f}}_{\bm{k}}$ is the free Hamiltonian of the scalar fields, $Z\equiv\mathrm{Tr}(\mathrm{e}^{-\beta H_{{F}}})=\prod\limits_{\bm{k}}(1-e^{-\beta\omega_{\bm{k}}})^{-1}$, and on the right of this equation, we have used the expansion of the free fields, Eq.~(\ref{free-field}).
After further simplifications, we find that there are only six nonvanishing terms left on the right of this equation, the coefficients of which are respectively given by
\bea
\left\{
  \begin{array}{ll}
\frac{1}{Z}\mathrm{Tr}(\mathrm{e}^{-\beta H_{{F}}}a^{\dag}_{\bm{k}_1}a^{\dag}_{\bm{k}_2}a_{\bm{k}_3}a_{\bm{k}_4})=
N_{\bm{k}_1}N_{\bm{k}_2}(\delta_{\bm{k}_1\bm{k}_3}\delta_{\bm{k}_2\bm{k}_4}+\delta_{\bm{k}_1\bm{k}_4}\delta_{\bm{k}_2\bm{k}_3})\;,\\
\frac{1}{Z}\mathrm{Tr}(\mathrm{e}^{-\beta H_{{F}}}a^{\dag}_{\bm{k}_1}a_{\bm{k}_2}a^{\dag}_{\bm{k}_3}a_{\bm{k}_4})=
N_{\bm{k}_1}N_{\bm{k}_3}\delta_{\bm{k}_1\bm{k}_2}\delta_{\bm{k}_3\bm{k}_4}+N_{\bm{k}_1}(N_{\bm{k}_3}+1)\delta_{\bm{k}_1\bm{k}_4}\delta_{\bm{k}_2\bm{k}_3}\;,\\
\frac{1}{Z}\mathrm{Tr}(\mathrm{e}^{-\beta H_{{F}}}a^{\dag}_{\bm{k}_1}a_{\bm{k}_2}a_{\bm{k}_3}a^{\dag}_{\bm{k}_4})=
N_{\bm{k}_1}(N_{\bm{k}_4}+1)(\delta_{\bm{k}_1\bm{k}_2}\delta_{\bm{k}_3\bm{k}_4}+\delta_{\bm{k}_1\bm{k}_3}\delta_{\bm{k}_2\bm{k}_4})\;,\\
\frac{1}{Z}\mathrm{Tr}(\mathrm{e}^{-\beta H_{{F}}}a_{\bm{k}_1}a^{\dag}_{\bm{k}_2}a^{\dag}_{\bm{k}_3}a_{\bm{k}_4})=
N_{\bm{k}_2}(N_{\bm{k}_3}+1)\delta_{\bm{k}_1\bm{k}_3}\delta_{\bm{k}_2\bm{k}_4}+(N_{\bm{k}_2}+1)N_{\bm{k}_3}\delta_{\bm{k}_1\bm{k}_2}\delta_{\bm{k}_3\bm{k}_4}\;,\\
\frac{1}{Z}\mathrm{Tr}(\mathrm{e}^{-\beta H_{{F}}}a_{\bm{k}_1}a^{\dag}_{\bm{k}_2}a_{\bm{k}_3}a^{\dag}_{\bm{k}_4})=
(N_{\bm{k}_2}N_{\bm{k}_4}+N_{\bm{k}_2}+N_{\bm{k}_4}+1)\delta_{\bm{k}_1\bm{k}_2}\delta_{\bm{k}_3\bm{k}_4}
+N_{\bm{k}_2}(N_{\bm{k}_4}+1)\delta_{\bm{k}_1\bm{k}_4}\delta_{\bm{k}_2\bm{k}_3}\;,\\
\frac{1}{Z}\mathrm{Tr}(\mathrm{e}^{-\beta H_{{F}}}a_{\bm{k}_1}a_{\bm{k}_2}a^{\dag}_{\bm{k}_3}a^{\dag}_{\bm{k}_4})=
(N_{\bm{k}_3}N_{\bm{k}_4}+N_{\bm{k}_3}+N_{\bm{k}_4}+1)(\delta_{\bm{k}_1\bm{k}_3}\delta_{\bm{k}_2\bm{k}_4}+\delta_{\bm{k}_1\bm{k}_4}\delta_{\bm{k}_2\bm{k}_3})\;,
  \end{array}
\right.
\label{aaaa}
\eea
with $N_{\bm{k}_i}=(\mathrm{e}^{\beta \omega_{\bm{k}_i}}-1)^{-1}$. Hereafter, we omit the subscripts ``$f$" of the creation and annihilation operators for brevity.

Notice that for the two-point correlation function of the scalar fields in the same thermal state,
\bea
\langle \beta|\phi(x_1)\phi(x_2)|\beta\rangle&\equiv&\frac{1}{Z}\mathrm{Tr}\left[\mathrm{e}^{-\beta H_{{F}}}\phi(x_1)\phi(x_2)\right]\nonumber\\
&=&\frac{1}{Z}\sum_{\bm{k}_1,\bm{k}_2}\mathrm{Tr}\left\{\mathrm{e}^{-\beta H_{{F}}}\left[a_{\bm{k}_1}f_{\bm{k}_1}(x_1)+\mathrm{H. c.}\right]\left[a_{\bm{k}_2}f_{\bm{k}_2}(x_2)+\mathrm{H. c.}\right]\right\}\nonumber\\
&=&\sum_{\bm{k}}\left[N_{\bm{k}}f^*_{\bm{k}}(x_1)f_{\bm{k}}(x_2)+(N_{\bm{k}}+1)f_{\bm{k}}(x_1)f^*_{\bm{k}}(x_2)\right]\;.
\eea
Here to obtain the third line of the above equation, we have used the following relations:
\bea
\left\{
  \begin{array}{ll}
\mathrm{Tr}(\mathrm{e}^{-\beta H_{{F}}}a_{\bm{k}_1}a_{\bm{k}_2})=\mathrm{Tr}(\mathrm{e}^{-\beta H_{{F}}}a^{\dag}_{\bm{k}_1}a^{\dag}_{\bm{k}_2})=0\;,\\
\mathrm{Tr}(\mathrm{e}^{-\beta H_{{F}}}a^{\dag}_{\bm{k}_1}a_{\bm{k}_2})=\delta_{\bm{k}_1\bm{k}_2}\frac{1}{\mathrm{e}^{\beta \omega_{\bm{k}_1}}-1}\prod\limits_{\bm{k}}\frac{1}{1-\mathrm{e}^{-\beta\omega_{\bm{k}}}}\;,\\
\mathrm{Tr}(\mathrm{e}^{-\beta H_{{F}}}a_{\bm{k}_1}a^{\dag}_{\bm{k}_2})=\delta_{\bm{k}_1\bm{k}_2}\left(1+\frac{1}{\mathrm{e}^{\beta \omega_{\bm{k}_1}}-1}\right)\prod\limits_{\bm{k}}\frac{1}{1-\mathrm{e}^{-\beta\omega_{\bm{k}}}}\;.
  \end{array}
\right.
\label{aa}
\eea

Combining Eqs.~(\ref{aaaa})-(\ref{aa}) with Eq.~(\ref{fourpoint}), we can obtain
\bea
\langle \beta|\phi(x_1)\phi(x_2)\phi(x_3)\phi(x_4)|\beta\rangle&=&\langle \beta|\phi(x_1)\phi(x_2)|\beta\rangle\langle \beta|\phi(x_3)\phi(x_4)|\beta\rangle\nonumber\\
&&+\langle \beta|\phi(x_1)\phi(x_3)|\beta\rangle\langle \beta|\phi(x_2)\phi(x_4)|\beta\rangle\nonumber\\
&&+\langle \beta|\phi(x_1)\phi(x_4)|\beta\rangle\langle \beta|\phi(x_2)\phi(x_3)|\beta\rangle\;.
\eea
Then use this relation in Eqs.~(\ref{1tf}) and (\ref{1rr}), and take the expectation values of them over the ground state $|g_{{A}}g_{{B}}\rangle$, and we finally get Eqs.~(\ref{2tf}) and (\ref{2rr}).

\end{document}